\documentclass[letter,twocolumn]{jpsj2}

\usepackage{amsmath}
\usepackage{graphicx}

\newcommand{\pade}{Pad\'e }

\title{Development of superconducting correlation at low temperatures in the two-dimensional $t$-$J$ model}

\author{\textsc{Takashi Koretsune}$^{1}$\thanks{E-mail address: koretune@riken.jp} and \textsc{Masao Ogata}$^{2}$}
\inst{$^{1}$RIKEN, 2-1 Hirosawa, Wako, Saitama 351-0198, Japan\\
$^{2}$Department of Physics, University of Tokyo, Bunkyo-ku, Tokyo 113-0033, Japan}

\abst{
The equal-time pairing correlation function of the two-dimensional $t$-$J$ model 
on a square lattice is studied using a high-temperature expansion method.
The sum of the pairing correlation, its spatial dependence, and
the correlation length are obtained as functions of temperature
down to $T \simeq 0.2 t$.
By comparison of single-particle contributions in the correlation functions,
we find an effective attractive interaction between quasi-particles
in $d_{x^2-y^2}$-wave pairings.
It is shown that $d$-wave correlation grows rapidly at low temperatures for
$0.5 < n < 0.9$, with $n$ being the electron density.
The temperature for this growth is roughly scaled by $J/2$.
This is in sharp contrast to the Hubbard model in a weak
or intermediate coupling region,
where there is no numerical evidence of superconductivity.
}

\kword{$t$-$J$ model, superconductivity}

\begin{document}
\maketitle

Since the discovery of high-$T_c$ superconductors\cite{bednorz1986},
great efforts have been devoted to explaining
the mechanism of their superconductivity.
It is now believed that two dimensionality and
strong electron correlation are important for
high-$T_c$ superconductivity.
Although many studies have been carried out using the
Hubbard model and the $t$-$J$ model on a two-dimensional
square lattice,
numerical studies using these models
do not always support superconductivity.
In the Hubbard model,
quantum Monte Carlo simulations show that there is no
enhancement of pairing correlation\cite{imada1989,furukawa1992,moreo1992,zhang1997},
although RPA\cite{scalapino1995}, fluctuation exchange (FLEX) approximations\cite{bickers1989} or
renormalization group approaches\cite{halboth2000}
claim $d_{x^2-y^2}$-wave superconductivity.
It is considered that the lattice size or electron correlation $(U/t < 4)$ 
in the numerical calculation is too small to
deal with superconductivity which has a very low energy scale\cite{yokoyama2004}.
Although several efforts to detect superconductivity have been made\cite{kuroki1997},
it is far from conclusive whether superconductivity occurs
in the two-dimensional Hubbard model numerically.

In the $t$-$J$ model, on the other hand,
mean-field theory indicates $d_{x^2-y^2}$-wave superconductivity
and superconductivity is supported by several numerical calculations
\cite{kohno1997,dagotto1993-1,dagotto1993-2,yokoyama_shiba,gros1989,yokoyama_ogata,himeda_ogata,sorella2002,dagotto1994}.
However, these studies are restricted to small system sizes or a variational theory in the ground state.
In exact diagonalization studies of small clusters\cite{dagotto1993-1,dagotto1993-2},
it is very difficult to discuss the long-range order.
In variational calculations\cite{yokoyama_shiba,gros1989,yokoyama_ogata,himeda_ogata,sorella2002},
on the other hand, it is known that the variational energy is mainly determined
from short-range correlations,
and thus it is difficult to confirm the existence of superconducting long-range order.
Therefore, some exact numerical calculations are highly necessary.
Since quantum Monte Carlo simulations have not been successful in the $t$-$J$ model,
the only available exact calculation is a high-temperature expansion method,
which has no negative sign problem and can be used to treat large systems exactly.
Using this method, we can study the properties at finite temperatures
which give complementary information for ground-state studies.

Using high-temperature expansions, thermodynamic quantities, and two-point correlation functions,
such as free energy and spin-spin correlation functions, have been studied\cite{putikka92,
singh92,putikka94}.
However, superconducting pairing correlations have not been obtained
because calculation of these quantities, which are essentially four-point correlations,
requires a greater number of graphs.
Recently, Pryadko {\it et al.}\cite{pryadko2004} have obtained a high-temperature series of pairing susceptibilities through
the ninth order in inverse temperature and claimed that the pairing susceptibility does not grow at 
low temperatures for $J/t < 1$.
However, the order of the expansion is too low and
the error bars are too large to make a definite conclusion.

In this study, we investigate the high-temperature expansion
of equal-time pairing correlation
functions up to the twelfth order in inverse temperatures.
In fact, series analysis of a pairing correlation function
is a difficult problem since pairing correlation functions decay very rapidly for long distances as discussed below.
To overcome this difficulty,
we develop a method of observing superconductivity on the basis of correlation lengths.
In this analysis, we find that it is possible to study superconductivity more accurately.
Using \pade approximations,
we successfully extrapolate the series down to $T \sim 0.2t$ and find that
the pairing correlation for $d_{x^2-y^2}$-wave symmetry 
grows rapidly at low temperatures for $J/t \sim 0.4$ and $0.5 < n < 0.9$.
By this calculation, we can explicitly show the development of superconducting
correlations in the $t$-$J$ model at finite temperatures.

The $t$-$J$ model is defined as
\begin{align}
\mathcal{H} = 
-t \sum_{\langle i,j \rangle \sigma} P \left( c_{i \sigma}^{\;\dagger} c_{j \sigma} + h.c. \right) P
+ J \sum_{\langle i,j \rangle} {\bf S}_i \cdot {\bf S}_j,
\end{align}
where summations are over the nearest-neighbor pairs on a square lattice
and the projection operator $P$
eliminates doubly occupied sites.
For this model, we calculate the high-temperature series of 
the equal-time pairing correlation function given as
\begin{align}
	P_{\alpha}(i, j) =
	\langle \Delta_{\alpha}^{\dagger}({\bf r}_i) \Delta_{\alpha}({\bf r}_j)
	+ \Delta_{\alpha}({\bf r}_i) \Delta_{\alpha}^{\dagger}({\bf r}_j) \rangle.
\end{align}
Here, $\alpha$ denotes the symmetry of the pairing function, and
the order parameter $\Delta_{\alpha}({\bf r}_i)$ is defined by
\begin{align}
	\Delta_{\alpha}({\bf r}_i) = \sum_{l} f_{\alpha}(l)
	( c_{i \uparrow} c_{i+l \downarrow} 
	\pm c_{i+l \uparrow} c_{i \downarrow}),
\end{align}
with $f_{\alpha}(l)$ being the form factor of the pairing correlation
and ${\bf r}_i$ being the coordinate at site $i$.
Considering the nearest neighbor pairings, the form factors in the square
lattice are given by
\begin{align}
	f_s(l) = \frac{1}{2} [ \delta_{l_x, 0}(\delta_{l_y, 1} + \delta_{l_y, -1})
	+ \delta_{l_y, 0}(\delta_{l_x, 1} + \delta_{l_x, -1}) ] ,\\
	f_d(l) = \frac{1}{2} [ \delta_{l_x, 0}(\delta_{l_y, 1} + \delta_{l_y, -1})
	- \delta_{l_y, 0}(\delta_{l_x, 1} + \delta_{l_x, -1}) ] ,
\end{align}
for a singlet case, and
\begin{align}
	f_p(l) = \frac{1}{\sqrt{2}} [ \delta_{l_y, 0}(\delta_{l_x, 1} - \delta_{l_x, -1}) ],
\end{align}
for a triplet case.
Each denotes the pairing correlation for an
extended $s$-wave, $d_{x^2-y^2}$-wave and $p$-wave, respectively.
The average in eq.\ (2) is taken over the grand canonical ensemble; that is,
$\langle O \rangle = {\rm Tr} O e^{-\beta(\mathcal{H} - \mu N_e)}/
{\rm Tr} e^{-\beta(\mathcal{H} - \mu N_e)}$.
We calculate $P_{\alpha}(i, j)$ up to the twelfth order in inverse temperature, $\beta$,
using the finite cluster method\cite{dombgreen3}.
Since we consider Cooper pairs of nearest-neighbor sites, graphs appearing in
the calculation have up to fourteen edges.
In each graph, a pair of edges represents Cooper pairs at site $i$ and $j$,
and the rest of the edges represent
the corresponding nearest-neighbor bonds in the Hamiltonian.
After calculating $P_{\alpha}(i, j)$ as a function of $\beta$, $\lambda = e^{\beta \mu}$, $J$ and $t$,
we eliminate $\lambda$ by solving
the high-temperature series of electron density, $n$, given by
\[
	n = - \frac{\partial \Omega}{\partial \mu} = \frac{2 \lambda}{1+2\lambda} + \cdots,
\]
and obtain $P_{\alpha}(i, j)$ as a function of $n$\cite{kubo_tada2}.
The obtained series is too long to show here,
but it is available from the authors upon request.

For extrapolating a high temperature series to low temperatures,
we use a \pade approximation.
To improve the convergence, we also calculate the series using an alternative choice of
the variable $w(\beta) = \tanh(f \beta)$,
where $f$ is a parameter tuned to optimize the convergence\cite{singh92}.
For small values of $f$, the series does not differ much from that for $\beta$.
Note that when $\beta \rightarrow \infty$, $w$ becomes finite,
which avoids the unphysical divergences or zeros
of a \pade approximation at $\beta = \infty$.
However, one cannot extrapolate to values of $\beta$
much larger than $1/f$ since $\tanh(f \beta)$ almost saturates.
With this in mind, we generate various \pade approximations
using $0.01 < f < 0.2$.
The extrapolated values and the error bars of all the following figures
are calculated as averages and deviations of various \pade
approximations, respectively.

\begin{figure}
	\begin{center}
	\includegraphics[scale=0.65]{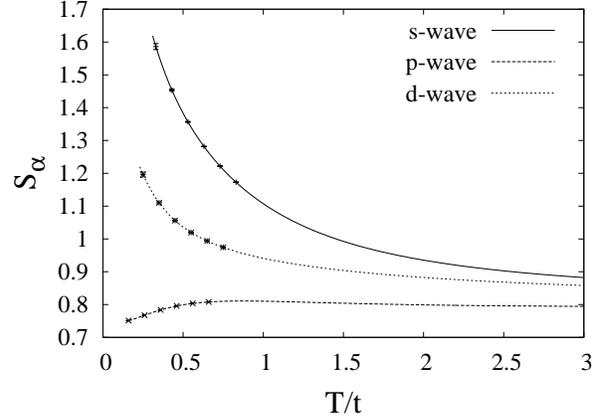}
	\end{center}
	\caption{
		Sum of equal-time pairing correlation functions,
		$S_{\alpha}(T)$, for $\alpha = s, p,$ and $d$ at
		$n = 0.8$ and $J/t = 0.4$.
	}
	\label{fig:susp_super}
\end{figure}

Let us first discuss the sum of the equal-time pairing correlation function (${\bf q}=0$
component), $S_{\alpha}(T)$, defined as
\begin{align}
	S_{\alpha} = \frac{1}{N} \sum_{i, j} P_{\alpha}(i, j).
\end{align}
Figure \ref{fig:susp_super} shows $S_{\alpha}(T)$ ($\alpha=s,p,$ and $d$-wave) at $J/t=0.4$ for $n=0.8$.
If superconductivity is realized at $T=0$, $S_{\alpha}(T)$ diverges as $T \rightarrow 0$
together with pairing susceptibility.
It can be seen in Fig.\ \ref{fig:susp_super}
that $p$-wave pairing correlation does not show an enhancement at low temperatures.
This is the expected behavior since the effect of $J$, which favors singlet pairing, is large 
near half filling.
In contrast, $s$- and $d$-wave pairing correlations grow at low temperatures ($T < 0.5t$).
In particular, $s$-wave pairing increases rapidly, which at first glance suggests an $s$-wave superconductivity.
However, as we show shortly, we find that this $s$-wave enhancement is due to a local singlet formation.

\begin{figure}
	\begin{center}
	\includegraphics[scale=0.65]{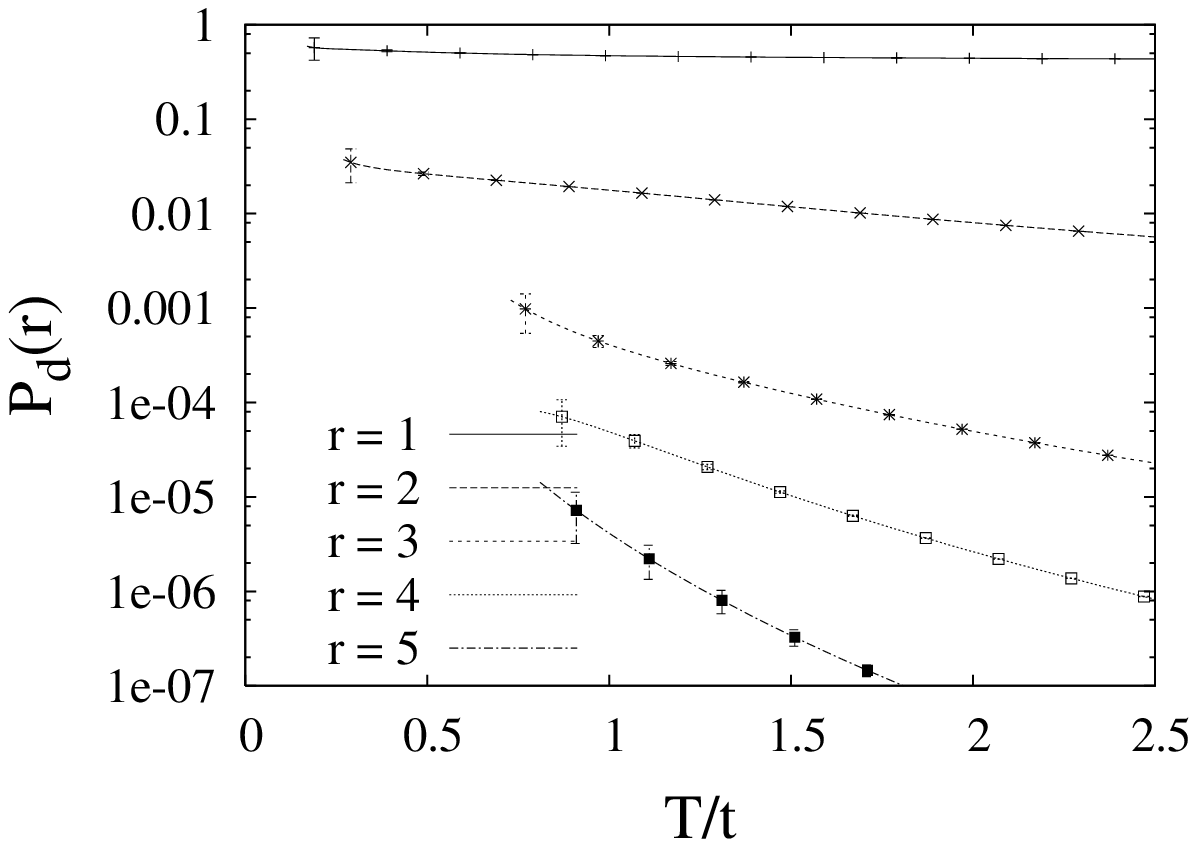}\\
	\includegraphics[scale=0.65]{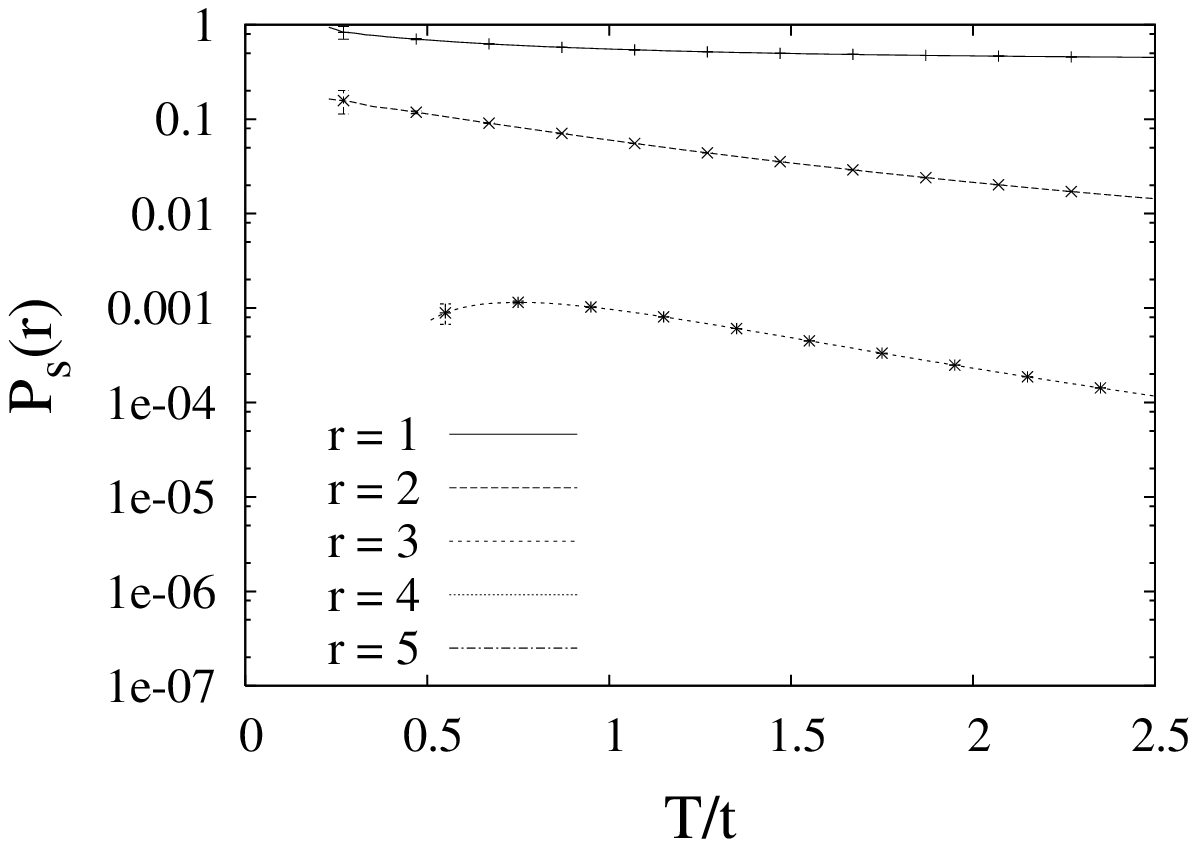}
	\end{center}
	\caption{
		Distance dependence of pairing correlation function, $P_{\alpha}(r)$, at $J/t = 0.4, n = 0.75$
		for (i) $\alpha = d$ and (ii) $\alpha = s$.
		In the case of $d$-wave, $P_{d}(r)$ with $r=1 \sim 5$ grow with decreasing temperatures
		though absolute values are small for long distances.
		In contrast, $P_s(r)$ with $r>3$ are not found in this figure
		since $P_s(r)$ at $r=4$ or $5$ is negative in this case.
	}
	\label{fig:corr_distance}
\end{figure}

The pairing correlations as functions of distance, $P_{\alpha}(r)$, given by
\begin{align}
	P_{\alpha}(r) = \frac{1}{N} \sum_{|{\bf r}_i-{\bf r}_j| = r} P_{\alpha}(i, j),
\end{align}
with $\alpha = s$ or $d$ are shown in Fig.\ \ref{fig:corr_distance}.
Here, the distance is defined as $|{\bf r}| = |r_x| + |r_y|$.
In agreement with the behavior of $S_d$, $P_d(r)$ grows with decreasing temperature.
Figure \ref{fig:corr_distance}(a) shows that
$P_d(r)$ with $r>2$ will reach values of $0.01$ or larger as $T \rightarrow 0$,
indicating that the long-range pairing correlation exists at $T = 0$.
In contrast, the $s$-wave pairing correlation becomes very small
(or even negative in some parameters) for long distances ($r>3$).
This means that the enhancement of $S_s$ in Fig.\ \ref{fig:susp_super}
comes only from the short-range pairing correlation and that
$S_s$ does not diverge as $T \rightarrow 0$.
These results are consistent with the ground state studies.

\begin{figure}
	\begin{center}
		\includegraphics[scale=0.65]{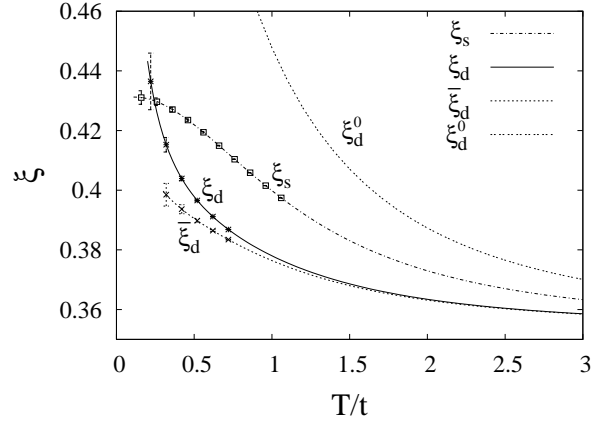}
	\end{center}
	\caption{
		Correlation lengths for $s$- and $d$-wave pairing at $J/t = 0.4$ and $n = 0.8$.
		For comparison, the correlation length of $d$-wave pairing for the free-electron system, $\xi_d^0$,
		and the correlation length of single-particle contribution (see the text), $\overline{\xi_d}$,
		are also shown.
		}
	\label{fig:corr_length}
\end{figure}

Although the result of $P_{\alpha}(r)$ shows an enhancement of the long-range $d$-wave pairing correlation,
the extrapolation of $P_{\alpha}(r)$ can be done only at relatively high temperatures $(T/t > 1)$
for large $r$ values, where the high-temperature series is rather short.
For example, the series of $P_{\alpha}(r)$ at $r=5$ 
starts from the order of $\beta^8$.
Thus, an alternative method is desired to clarify the growth of long-range correlation
at low temperatures more accurately.
For this purpose, 
we calculate the correlation length of pairing correlations as follows.
Let us consider the Fourier transform of $P_{\alpha}(i, j)$,
\begin{align}
      P_{\alpha}({\bf q}) = \frac{1}{N}\sum_{i, j} P_{\alpha}(i, j) \exp(i {\bf q} \cdot ({\bf r}_i - {\bf r}_j) ).
\end{align}
When the correlation of $\Delta_{\alpha}$ develops,
$P_{\alpha}({\bf q})$ shows a peak at $q = 0$.
Then, the correlation length of $\Delta_{\alpha}$ is defined from the width of its peak as
\begin{align}
	\xi^2_{\alpha}
	&= \left. -\frac{1}{2 d P_{\alpha}({\bf q})} \frac{\partial^2}{\partial {\bf q}^2} P_{\alpha}({\bf q}) \right|_{q = 0}\nonumber \\
	&= \frac{1}{2 d S_{\alpha}} \sum_{i} |{\bf r}_i|^2 \langle \Delta_{\alpha}^\dagger(0) \Delta_{\alpha}({\bf r}_i) + \Delta_{\alpha}(0) \Delta_{\alpha}^\dagger({\bf r}_i) \rangle,
\end{align}
with $d$ being a dimension.
We find that the high-temperature series of this quantity converges
much better than $P_{\alpha}(r)$.
Figure \ref{fig:corr_length} shows $\xi_s$ and $\xi_d$ at $n = 0.8$.
It is apparent that $\xi_d$
shows a large enhancement for $T<0.1$-$0.2t$,
which is consistent with the behavior of $S_d$ and $P_d(r)$.
In contrast, $\xi_s$ almost saturates at $T \sim 0.4t$ 
and does not grow at low temperatures, which also agrees
with the behavior of $P_s(r)$.

Although the enhancement of $\xi_d$ is promising for superconducting long-range
order at $T=0$,
we must be careful with our conclusion since the pairing correlation contains
single-particle contributions.
For example, the correlation length of $d$-wave pairing symmetry
in the free-electron system, $\xi_d^0$, for the same electron density ($n=0.8$)
is shown in Fig.\ \ref{fig:corr_length}.
Although $\xi_d^0$ is relatively large and grows as $T\rightarrow0$,
there should be no instability of superconductivity.
This enhancement of $\xi_d^0$
comes from the single-particle part:
for free systems,
$\langle c_{i \uparrow} c_{j \downarrow} c_{k \downarrow}^{\dagger} c_{l \uparrow}^{\dagger} \rangle =
\langle c_{i \uparrow} c_{l \uparrow}^{\dagger} \rangle 
\langle c_{j \downarrow} c_{k \downarrow}^{\dagger} \rangle$
holds, and the single-particle part 
$\langle c_{i \uparrow} c_{l \uparrow}^{\dagger} \rangle$ provides the enhancement.

To extract an intrinsic pairing correlation,
we use a method employed in the Hubbard model\cite{white1989}.
The single-particle contribution in $P_\alpha(i,j)$ can be obtained
from high-temperature series of 
$\langle c_{i \uparrow} c_{l \uparrow}^{\dagger} \rangle$ and
$\langle c_{j \downarrow} c_{k \downarrow}^{\dagger} \rangle$.
(They correspond to disconnected diagrams.)
By applying the same extrapolation scheme used for $\xi_d$,
we calculate $\overline{\xi_d}$ from
$\langle c_{i \uparrow} c_{l \uparrow}^{\dagger} \rangle 
\langle c_{j \downarrow} c_{k \downarrow}^{\dagger} \rangle$.
Then, comparing $\xi_d$ with $\overline{\xi_d}$,
we can discuss the effective interactions among pairs.
The obtained results using the $t$-$J$ model are shown in Fig.\ \ref{fig:corr_length}.
It is observed that $\overline{\xi_d}$ is strongly suppressed compared with $\xi_d^0$,
which indicates that electrons have difficulty moving with strong correlations.
Furthermore, $\xi_d$, which is very similar to $\overline{\xi_d}$ at high temperatures,
deviates from $\overline{\xi_d}$ and starts to grow at low temperatures.
This means that there is an effective attractive interaction 
in the $d$-wave pairings.
The growth of correlation length itself means that the pairing correlation becomes
long-ranged with lowering of the temperature.
In fact, the behavior of $\xi_d$ is apparently different from that of the $s$-wave case, $\xi_s$,
where pairing correlations are short-ranged.
Although the divergence of $\xi_d$ as $T\rightarrow0$ is not perfectly proved,
the behavior of $\xi_d$ at $T < 0.3t$ strongly suggests
superconductivity in the ground state.

\begin{figure}[thp]
	\begin{center}
	\includegraphics[scale=0.65]{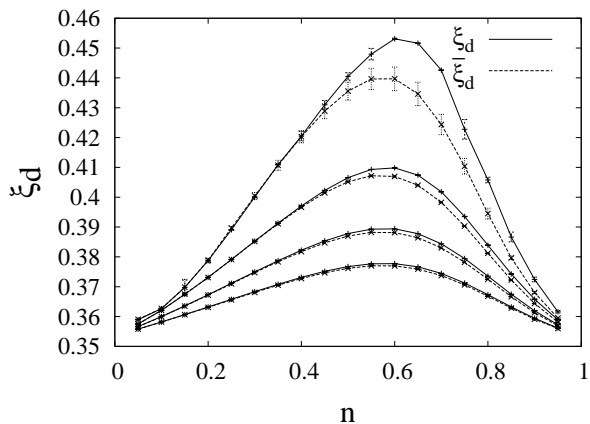}
	\end{center}
	\caption{
		$\xi_d$ and $\overline{\xi_d}$ as functions of $n$ at $T/t = 0.4, 0.8, 1.2,$ and $1.6$
		(from top to bottom).
	}
	\label{fig:phase}
\end{figure}

The doping rate dependences of $\xi_d$ and $\overline{\xi_d}$ are shown in Fig.\ \ref{fig:phase}
for some values of $T/t$.
Apparently, $\xi_d$ shows enhancement over a wide region.
However, in the low-density region, $\xi_d$ and $\overline{\xi_d}$ are very similar
since the effects of $J$ and the Gutzwiller projection,
which makes a difference between $\xi_d$ and $\overline{\xi_d}$, are small.
We can see that $\xi_d$ grows faster than $\overline{\xi_d}$ in the range $0.5 < n < 0.9$.
Thus, superconductivity is expected in this region.
This doping dependence is consistent with that observed in
previous ground state studies\cite{kohno1997,yokoyama_shiba,himeda_ogata}.

Here, we describe the $J$ dependence of pairing correlations.
It is found that the behavior of $\xi_d$ for $J=0 \sim 1$ does not change qualitatively.
Even at $J=0$, $\xi_d$ grows with lowering temperatures, though it does not
show rapid growth at $T>0.1 \sim 0.2t$.
With increasing $J$, the growth of $\xi_d$ starts at higher temperatures.
It is roughly scaled by $J/2$ at $n = 0.8$.

To summarize, we have studied the equal-time pairing correlation function
of the two-dimensional $t$-$J$ model on a square lattice.
It is found that the pairing correlation grows with decreasing temperature.
A prominent enhancement of correlation length is detected in a wide region,
which occurs at $T<0.2t$.
This growth of the pairing correlation length
indicates the superconducting ground state in the $t$-$J$ model.
This is in sharp contrast to the Hubbard model in a weak or intermediate
coupling region.
Although the pairing susceptibility in the Hubbard model shows an
effective attractive interaction in the $d$-wave channel\cite{white1989},
pairing correlations are short-ranged
and do not show scaling with lattice size\cite{furukawa1992,moreo1992}.


\begin{thebibliography}{99}

\bibitem{bednorz1986}
J.~Bednorz and K.~M\"uller: Z. Phys. B \textbf{64} (1986) 189.

\bibitem{imada1989}
M.~Imada and Y.~Hatsugai: J. Phys. Soc. Jpn. \textbf{58} (1989) 3752.

\bibitem{furukawa1992}
N.~Furukawa and M.~Imada: J. Phys. Soc. Jpn. \textbf{61} (1992) 3331.

\bibitem{moreo1992}
A.~Moreo: Phys. Rev. B \textbf{45} (1992) 5059.

\bibitem{zhang1997}
S.~Zhang, J.~Carlson and J.~E. Gubernatis: Phys. Rev. Lett. \textbf{78} (1997)
  4486.

\bibitem{scalapino1995}
D.~J. Scalapino: Phys. Rep. \textbf{250} (1995) 329.

\bibitem{bickers1989}
N.~E. Bickers, D.~J. Scalapino and S.~R. White: Phys. Rev. Lett. \textbf{62}
  (1989) 961.

\bibitem{halboth2000}
C.~J. Halboth and W.~Metzner: Phys. Rev. B \textbf{61} (2000) 7364.

\bibitem{yokoyama2004}
H.~Yokoyama, Y.~Tanaka, M.~Ogata and H.~Tsuchiura: J. Phys. Soc. Jpn.
  \textbf{73} (2004) 1119.

\bibitem{kuroki1997}
K.~Kuroki and H.~Aoki: Phys. Rev. B \textbf{56} (1997) R14287.

\bibitem{kohno1997}
M.~Kohno: Phys. Rev. B \textbf{55} (1997) 1435.

\bibitem{dagotto1993-1}
E.~Dagotto and J.~Riera: Phys. Rev. Lett. \textbf{70} (1993) 682.

\bibitem{dagotto1993-2}
E.~Dagotto, J. Riera, Y.~C. Chen, A. Moreo, A. Nazarenko, F. Alcaraz and F. Ortolani: Phys. Rev. B \textbf{49} (1994) 3548.

\bibitem{yokoyama_shiba}
H.~Yokoyama and H.~Shiba: J. Phys. Soc. Jpn. \textbf{57} (1988) 2482.

\bibitem{gros1989}
C.~Gros: Ann. Phys. (N.Y.) \textbf{189} (1989) 53.

\bibitem{yokoyama_ogata}
H.~Yokoyama and M.~Ogata: J. Phys. Soc. Jpn. \textbf{65} (1996) 3615.

\bibitem{himeda_ogata}
A.~Himeda and M.~Ogata: Phys. Rev. B \textbf{60} (1999) R9935.

\bibitem{sorella2002}
S.~Sorella, G.~B. Martins, F.~Becca, C.~Gazza, L.~Capriotti, A.~Parola and
  E.~Dagotto: Phys. Rev. Lett. \textbf{88} (2002) 117002.

\bibitem{dagotto1994}
E.~Dagotto: Rev. Mod. Phys. \textbf{66} (1994) 763.

\bibitem{putikka92}
W.~O. Putikka, M.~U. Luchini and T.~M. Rice: Phys. Rev. Lett. \textbf{68}
  (1992) 538.

\bibitem{singh92}
R.~R.~P. Singh and R.~L. Glenister: Phys. Rev. B \textbf{46} (1992) 11871.

\bibitem{putikka94}
W.~O. Putikka, R.~L. Glenister, R.~R.~P. Singh and H.~Tsunetsugu: Phys. Rev.
  Lett. \textbf{73} (1994) 170.

\bibitem{pryadko2004}
L.~P. Pryadko, S.~A. Kivelson and O.~Zachar: Phys. Rev. Lett. \textbf{92}
  (2004) 067002.

\bibitem{dombgreen3}
C.~Domb and M.~S. Green: \emph{Phase Transitions and Critical Phenomena},
  vol.~3 (Academic, London, 1974).

\bibitem{kubo_tada2}
K.~Kubo and M.~Tada: Prog. Theor. Phys. \textbf{71} (1984) 479.

\bibitem{white1989}
S.~R. White, D.~J. Scalapino, R.~L. Sugar, N.~E. Bickers and R.~T. Scalettar:
  Phys. Rev. B \textbf{39} (1989) R839.

\end{thebibliography}
\end{document}